\documentstyle[preprint,aps,epsf]{revtex}
\begin{document}
\preprint{IASSNS-HEP-97/87}
\draft
\title{The Quantum-Classical Metal}
\author{David G. Clarke$^1$, S. P. Strong$^2$, P. M. Chaikin$^1$ and
E. I. Chashechkina$^1$
}
\address{$^1$Joseph Henry Laboratories of Physics,
Princeton University, P.O. Box 708, Princeton, NJ 08544 \\
$^2$Institute for Advanced Study, Olden Lane, Princeton, NJ 08540, USA.\\}
\date{July 16, 1997}
%\date{\today}
\maketitle
\begin{abstract}

In a normal Fermi liquid,
Landau's theory precludes the loss of 
single fermion, quantum coherence in the low energy/temperature limit.
For highly anisotropic, strongly correlated metals
there is no proof that this remains
the case: we propose that quantum coherence for transport 
in some 
directions may be lost intrinsically.
This should stabilize a novel,
qualitatively anisotropic non-Fermi
liquid, separated by a novel
zero temperature, quantum phase transition from the
Fermi liquid state and categorized by the unobservability
of certain interference effects.
There is compelling experimental evidence for 
this transition as a function of magnetic field 
in the metallic phase of the organic conductor (TMTSF)$_2$PF$_6$.

\end{abstract}

%\pacs{PACS numbers: 
%71.27.+a, 72.10.-d 
%74.25.Fy, 
%74.72.-h
%}

%\begin{multicols}{2}

%\narrowtext

\newpage

\section{Introduction}
For decades, the accepted theory of metallic states
of electronic matter has been Landau's Fermi liquid theory (FLT).
Recently, in the context of systems
with strong electronic correlations, the universality of
FLT has come into question.
It is generally agreed that correlations are most dangerous
to FLT when the electrons live in low ({\em i.e.\/} one or two) dimensions,
and thus recent interest in breakdowns of FLT has focused on systems,
such as the cuprate and organic superconductors
(above their respective superconducting transition temperatures or in strong
magnetic fields), which are ``metallic'', bulk crystals, but 
enormously anisotropic.  
In these systems, the bare Coulomb interaction
is not small, and as such it is
not {\em a priori\/} obvious that it can be dealt with by perturbing
about an anisotropic three-dimensional (3D) Fermi gas, as FLT requires.  
It is more natural to
think of such systems as collections of strongly correlated
one or two dimensional electron liquids with weak, interliquid single particle
hopping.
Since an essential characteristic of a Fermi liquid is the
coherence of single electron motion in {\em all\/} spatial directions,
the question arises as to whether such coherence is guaranteed to
occur in the presence of anisotropy and strong interaction.
The purpose of this paper is to introduce the reader to both theoretical
and experimental considerations strongly challenging 
the universality and inevitability of such coherence. In fact, we propose
here the existence of a new metallic, non-Fermi liquid state in which 
3D quantum coherence is intrinsically destroyed by strong correlations
{\em at zero temperature and in a pure crystal}.
To orient the reader, we begin
with a discussion of quantum coherence and how such coherence 
can be lost, leading to incoherent, or ``classical'', behavior.

\section{Decoherence and the Quantum-Classical ``Boundary''}
According to the Copenhagen interpretation of
quantum mechanics there exists a remarkable,
qualitative distinction
between the ``microscopic'' and ``macroscopic''
realms.
The microscopic world is governed by quantum mechanics, the
macroscopic world by classical mechanics; ``observers'' and
``measuring apparati'' {\em etc.\/} lie in the latter, and 
``measurement'' and  the ``collapse of the wavefunction''
provide the procedure for describing 
the interactions of the micro- and macro-worlds. 
However, within the Copenhagen interpretation,
no explicit means of determining the scale at which things
become classical is given, and the means by which
classical behavior emerges at the macroscale is
not considered: it may or may not require new physics
beyond the quantum rules that govern the micro-world.
There is, however, no experimental evidence 
for a breakdown of quantum mechanics on any scale \cite{wheeler};
rather, classical behavior emerges in situations
where either (1) the system is too complex for the
quantum theory to make testable predictions,
or (2) the chief hallmark of classical behavior -
namely, the unobservability
of interference effects between different histories -
is in agreement with the predictions of quantum
theory {\it when one considers the effects of
the ``environment'' (the
unobserved degrees of freedom in the experiment) 
on the magnitude of these interference effects\/}.

Recall that, in the Schr\"{o}dinger picture of quantum mechanics, if a 
system is prepared in an energy eigenstate $\Psi_E$, then under 
time evolution the state picks up a temporal phase $-E\,t/\hbar$,
{\em i.e.\/}, $\Psi_E \rightarrow e^{-iEt/\hbar} \Psi_E$.
If a system is prepared in a superposition of 
such states,
then under time evolution each eigenstate develops a phase
independently of the others. If a subsequently applied
perturbation allows the mixing of these
states at a later time, then it is {\em in principle\/}
possible to observe the spectacular oscillation effects of quantum
mechanics. 
One of the more celebrated of these effects is strangeness
oscillations in Kaon decay, but it is also perfectly correct to
consider the ballistic motion of an electron wavepacket 
in an ordinary metal, as a true manifestation
of coherent interference effects.  
In these cases, the interference exists because the 
energies of the states mixed by the relevant ``perturbations''
(the weak interaction and the intersite hopping, respectively) are very narrowly
distributed with respect to the Hamiltonians in the absence of the
perturbations.  If one dealt instead with states which had energies specified
only to within some range $\Delta E$ (and the eigenstates of the system
are dense within this range) then the phase information would be 
lost in a time, $\Delta t \sim \hbar/\Delta E$, and
Kaon oscillations or any other quantum
interference effects
would only be observable on time scales shorter than or
comparable to $\Delta t$.
If the perturbation considered is too weak to have significant
effects on this time scale, then no interference effects
are observed and, further, discussing 
phase coherent superpositions of the mixed states is meaningless
because the dephasing eliminates all
the potentially observable coherence effects.

According to the decohering-histories interpretation
of quantum mechanics \cite{wheeler},
this loss of phase coherence is behind our generic inability to
observe quantum coherence effects in the macro-world:
macroscopic systems do obey quantum
mechanics; however, in a macroscopic system
it is generically a hopeless exercise to 
(1) specify all of the necessary degrees of freedom
to restrict the system to a sufficiently narrow range of 
initial energies,
and/or (2) find a perturbation which
connects an initial state only to states which are
themselves sufficiently degenerate in energy
for observing quantum effects.
In general, one can only restrict the wavefunction to a
manifold in Hilbert space and experimentally realizable
perturbations connect each state on this manifold to states
which lie on some other manifold in Hilbert space.
The energy of the states on these manifolds, {\em i.e.\/} the
energy of the ``initial'' and/or ``final'' states, is defined
only to within some spread $\Delta E$, and rapid dephasing is the 
result. From this point of view,
the Schr\"{o}dinger's cat {\em gedanken\/} experiment 
is spurious: it is meaningless to form superpositions
of $|{\rm dead}\rangle$ and $|{\rm alive}\rangle$ states of the cat,
since one cannot place the cat (even approximately)
in an initial energy eigenstate
nor find a perturbation which transforms the cat from an
``alive'' eigenstate to a group of sufficiently degenerate
``dead'' eigenstates to perform any kind of interference experiment.
It may be unsettling that within the Copenhagen
interpretation there is no deterministic way to know whether
the cat will be alive or dead upon opening the box, but 
the more unsettling
concept of the cat being in a superposition of 
$|{\rm dead}\rangle$ and $|{\rm alive}\rangle$ is meaningless
in the sense of having no observable consequences \cite{recentcat}.

\section{Decoherence in ``Strange'' Metals}
Interestingly,  the modern theory of clean metals, Fermi liquid theory
\cite{landau}, is an exception to many of the generic
statements made above:  it predicts observable interference effects
for certain {\em macroscopic\/} quantities.  This is of
particular interest because of the experimental evidence now
accumulating for the existence of ``strange'' metals, {\it i.e.\/}
materials whose properties are metallic, but which disobey various
predictions of FLT.  In this case, one might expect that 
some or all macroscopic quantum coherence which occurs for Fermi liquids
might be removed for these strange metals.  We will argue
here that this is the case, that there exist strange metallic
states in which the transport
in one or more directions 
is ``incoherent''  in the sense that
interference effects between histories which involve the transport
of electrons in such direction(s)
are unobservable {\em even in principle\/}
(in the low energy limit for pure systems).
This loss of coherence has strong physical consequences,
which we will argue 
{\em have been experimentally observed, \/}
demonstrating the existence of a new state of matter which might
best be described as a ``quantum-classical metal'' because of the 
lack of quantum coherence in some direction(s).

Such a state is intrinsically and qualitatively
anisotropic and could only be expected to arise 
in a material with strong anisotropy (and strong
electron-electron interactions).  In
fact, our proposal arose out of 
a careful study of the problem of one dimensional chains
of strongly interacting electrons, coupled by
a very weak interchain single electron hopping, $t_{\perp}$.   
It is useful for orientation to briefly describe that study.

We begin with
the observation that for short ranged, repulsive interactions,
electrons confined to one dimension form not
Fermi liquids but rather {\em Luttinger liquids\/} \cite{duncan},
a specific kind of ``non-Fermi liquid''.
The low energy properties of Luttinger liquids are well understood
and there is therefore a firm basis for studying the action of 
$t_{\perp}$ perturbatively 
%\cite{others,usprl,usadv}.
\cite{usprl,usadv}.
%Various studies based on this starting point
%\cite{others} have concluded that
It has long been known that
$t_{\perp}$ is the most relevant inter-liquid operator for a range of
interaction strengths, implying that for such
interactions, the correct low energy description
of the system requires that $t_{\perp}$
be retained regardless of how small it is. 
However, 
various earlier studies  
of the problem also either suggested or assumed implicitly that 
the leading relevance of $t_{\perp}$ was sufficient
to establish the existence of {\em coherent,\/} interchain
motion. The proposition is not self-evident
and it was an attempt to critically address it
for the coupled chains problem
\cite{usprl,usadv} that led to our  proposal 
of the quantum-classical metal.

For coupled Luttinger liquids,
there are a range of possible probes
of interchain interference effects
including, but not limited to, the
frequency dependent transverse conductivity,
the shape, if any, of the Fermi surface,
the tranverse bandwidth, the single particle
Green's function, {\em etc.\/} In the quantum-classical metal,
supposed to occur for sufficiently strong intrachain
interactions, all of these probes of interliquid
coherence should demonstrate the 
unobservability of interference effects between different histories
involving interchain hops.
Our original choice
for a natural probe was provided by the following
quantum oscillation effect: consider two identical chains
of strongly interacting electrons,
both in their isolated chain groundstates
at time $t=0$, but with chain 1
having $\delta N$ more electrons than
chain 2. At time $t=0$, turn on the interchain
hopping, $t_{\perp}$, and study the behavior
of $\langle \delta N(t) \rangle$. 
 This choice is motivated by the close connection
between $\langle \delta N(t) \rangle$
and a quantity, $\langle \sigma^z(t) \rangle$ 
\cite{usprl,usadv},
extensively studied in the 
simplest model for the 
quantum to classical crossover, the Caldeira-Leggett
model (hereafter the CL model) \cite{TLS}.

\section{
Electron hopping between non-Fermi liquids: 
Connection to a Simple Model of the Quantum-Classical Transition}
In the CL model, $\sigma$ represents the two state
``macroscopic'' degree of freedom (DOF) which is coupled to
an environment of infinitely many microscopic degrees
of freedom, represented by the simplest possible realization:
a bath of harmonic oscillators.  The oscillators ``measure''
the $\sigma^z$ state of the DOF, while an additional
perturbation (proportional to $\sigma^x$)
mixes the two $\sigma^z$ states coherently.
The coupling to the oscillators
should decohere the macroscopic degree of freedom
({\em i.e.\/} make superpositions of $\sigma^z$ states meaningless)
resulting in ``classical behavior'' for strong enough
coupling to the oscillators.  The quantity most frequently
studied is the expectation value, $\langle \sigma^z(t) \rangle$,
where the system has been prepared
by clamping $\sigma^z$ to $+1$ for all $t<0$ and allowing
the environment to adapt to this configuration. The
basic idea is that this resembles an experimentally
realizable situation:
the controllable, macroscopic
degree of freedom is held in a particular state and
the microscopic degrees of freedom, uncontrolled by the
experimenter, relax to their equilibrium under these
circumstances. The system is then released
and one looks for quantum interference effects
in the ensuing behavior of the observable,
macroscopic degree of freedom.  

One can also make a 
canonical transformation in the CL
problem, changing basis to the
eigenstates of the joint oscillator-DOF
system (in the absence of the $\sigma^x$
perturbation which mixes the two $\sigma^z$ eigenstates).
The CL Hamiltonian then takes the form
\[
H_{\rm CL} = \frac{1}{2} \Delta(\sigma^+ e^{-i \Omega}
+ ~{\rm h.c.}) +
\sum_i  \left(\frac{1}{2} m_i \omega_i
x_i^2  + \frac{1}{2m_i} p_i^2  \right) 
\]
where $C_i$ is the coupling to the $i$th 
harmonic oscillator,
$m_i$, $\omega_i$, $x_i$ and $p_i$
are the mass, frequency, position
and momentum of the $i$th oscillator,
and $\Omega = \sum_i \frac{C_i}{m_i \omega_i^2} p_i$.
In this language the ``measurement''
effects of the environment are encoded
in the {\em non-degeneracy\/} of the states
which are connected by the $\Delta$ term. 
In the absence of coupling to the bath ({\em i.e.\/} all $C_i$ zero),
the $\Delta$ term connects two {\em degenerate\/} $\sigma_z$ states.
In the presence of the bath, however, 
decoherence results when this degeneracy is sufficiently reduced
by the operator $e^{\pm i\Omega}$ which creates and destroys 
oscillator bosons over a broad energy range whenever a 
transition between the $\sigma^z$ states takes place.

In the new basis the usual CL preparation
amounts to taking the system to be in
one of the two groundstates of the system
in the absence of the $\Delta$ term,
and then suddenly switching this term on 
at time $t=0$. This is parallel to the preparation
discussed above for coupled Luttinger liquids;
$\sigma^z$ plays the role of $\delta N$,
$\Delta$ the role of $t_{\perp}$ and
the oscillator bath the role of the
well known charge and spin density
oscillator modes of the coupled Luttinger
liquids \cite{duncan}. 
In both cases one follows the dynamics
of the expectation value of
a discrete, observable ``macroscopic'' variable
(either $\sigma^z$ or $\delta N$) which has been
set up in a non-equilibrium state defined as
a valid groundstate of the problem 
in the absence of the perturbation
(either the $\sigma^x$ term or the inter-liquid
single particle hopping term).  As for the
$\Delta$ term in the TLS,
the action of
the single particle hopping can be written as
the product of
an operator whose only action is to change $\delta N$,
and an exponential in the creation and 
annihilation operators of the charge and spin density
oscillator modes of the coupled Luttinger
liquids, and
the resulting Hamiltonian is strikingly similar to the CL
Hamitonian, $H_{\rm CL}$ \cite{usadv}.
Despite the fact that the models cannot be mapped into
one another ({\em e.g.\/} the $\delta N$ variable is
many, not two, valued) the analogy does motivate the proposal
that incoherent dynamics can occur for
$\delta N$ in a similar manner to the
incoherence occuring for $\sigma^z$ \cite{usadv}.

The signature of
quantum coherence in the CL problem is 
taken to be the presence of
oscillations in $\langle \sigma^z(t) \rangle$,
which, when present, result from interference 
between histories in which $\sigma^z$ varies
differently.
We correspondingly take the presence of
oscillations in $\langle \delta N (t)\rangle$,
as the signature of quantum coherence
for the interchain hopping. 

In FLT,
$\langle \delta N \rangle$ exhibits oscillations
with frequency $Z t_{\perp}$ and damping which
vanishes in the limit of vanishing $t_{\perp}$ and
$\delta N(t=0)/L$ \cite{defs}.  
FLT therefore exhibits a dramatic instance of
macroscopic quantum coherence: not only
are the oscillations observable, they are essentially
undamped.   The source of this can
be traced back to the fact that, even though $\langle \delta N \rangle$
represents a macroscopic variable which is {\em a priori}
expected to couple to 
a huge number of uncontrolled (and potentially dephasing)
microscopic degrees of freedom, 
FLT dictates that $\langle \delta N \rangle$
is determined by the sum of many independent,
coherent, quasiparticle channels.  Since
each channel decouples from its environment 
in the limit of vanishing $\delta N(t=0)/L$,
coherence is unavoidable.
However,
for non-Fermi liquids, such as the Luttinger
liquids of the coupled chains problem,
there is no such special protection for
the coherence of $\langle \delta N \rangle$.

In fact, while FLT is analogous to the
CL model with no coupling between the
oscillators and $\sigma^z$,
coupled Luttinger liquids are analogous
to the CL model with finite coupling
to an ohmic bath of oscillators \cite{usprl,usadv}.
As a result, the most likely
behaviors for $\langle \delta N \rangle$
fall into three categories:
(1) for weak interactions, coherence and the
characteristic oscillations,
(2) for very strong interactions, localization
with $\langle \delta N(t \rightarrow \infty) \rangle \neq 0$
and (3) for a range of
intermediate interactions,
incoherence, with no oscillations in
$\langle \delta N \rangle$ but 
$\langle \delta N(t \rightarrow \infty) \rangle =  0$.
A remarkable result in the CL problem is that
the final possibility
is believed to occur over a broad range of coupling
constants, so that it is {\em not\/}
the case that the oscillations
simply become more heavily damped as the localization
behavior is approached.  Rather, the oscillation
frequency vanishes at some intermediate
coupling before the localization sets in.
The physical ingredients for
this behavior are also present in the coupled
Luttinger liquid problem \cite{usadv}
and we believe that
case (3) should occur there as well. 
If the oscillation frequency of 
$\langle \delta N \rangle$
is identically zero (over some range of
couplings where $t_{\perp}$ is the leading instability
of the uncoupled chains fixed point), then it is plausible that
{\em all\/} interchain interference effects are unobservable
in the low energy limit for this range of
couplings.  After all, there is
nothing special about $\langle \delta N \rangle$.
In the CL context, it is generally believed 
that the lack of coherence in
$\langle \sigma^z \rangle$ signals a general loss
of quantum coherence, and, in fact, the model was
intended in part to explain the absence of interference
effects for macroscopic objects. This absence is
manifestly generic, rather than being
limited to particular experiments. It is therefore
quite likely that the disappearance of
interference oscillations in $\langle \delta N \rangle$
represents a generic loss of coherence, rather than
one limited to the probe considered.  We correspondingly
expect that in this regime the transverse electrical
conductivity should lack a Drude peak, the single particle
Greens function should not exhibit a pole on the real axis 
which disperses with $k_{\perp}$ \cite{usadv}, no pair of split Fermi
surfaces should form, {\em etc.\/}

For two chains in the incoherent phase,
we expect there to be no  Fermi surface splitting,
which translates for infinitely many coupled chains into
the absence of warping of any higher dimensional Fermi surface.
This implies that, if it exists,
this regime constitutes a new state of matter
as the Fermi surface shape  gives a
clear, zero temperature, infinite time
distinction between the incoherent phase
and a normal metal.
We believe that the incoherent state 
is separated by a zero temperature 
quantum phase transition from a state where the 
interchain hopping is coherent and a three dimensional,
Fermi liquid metal occurs.  

At this point it is important to emphasize that our proposed
phase is {\em not\/} one in which the electrons are confined
to the chains (this would be the analog of (2) above for the
CL model). The latter is a phase in which, in the language of the 
renormalization group, $t_{\perp}$ is an {\em irrelevant\/}
operator. Our proposed phase is truly novel in that 
it is only the coherence which is confined, not the electrons - 
diffusive interchain motion still takes place, and 
$t_{\perp}$ is a {\em relevant\/} operator.

Our proposal is not limited
to the case of coupled Luttinger liquid chains.  While
theoretically this was the best controlled case to
study and the one most closely related to the
CL model, there are other possibilities.  In particular,
a set of strongly interacting two dimensional systems,
whose isolated groundstates were non-Fermi liquid metals
would be natural candidates for incoherent interplane
hopping.  It is this possibility which we will now
argue is experimentally realized in the organic
conductor (TMTSF)$_2$PF$_6$.

\section{The Quantum-Classical Metallic state in (TMTSF)$_2$PF$_6$}
(TMTSF)$_2$PF$_6$ is a Bechgaard salt 
composed of linear stacks of tetramethyltetraselenafulvalene
cations; the stacks are arranged into planes
separated by PF$_6$ anions which
provide overall charge neutrality and stabilize
the structure.  As expected from the structure,
the material is highly anisotropic with
resistive anisotropy at room temperature of
1:100:10$^5$ and possesses
a single, half-filled band \cite{dimer}.
(TMTSF)$_2$PF$_6$ is triclinic
so that the lattice vectors are not orthogonal
but roughly speaking
the $a$ axis lies along the stacks
(the most conducting direction), the $b$ axis
(the next most conducting direction)
in the TMTSF planes and the $c$ axis 
(the least conduction direction)
out of the planes.  At ambient pressure 
the material is a spin density wave insulator,
but at pressures above about 6 kilobar,
the ground state is superconducting.  It
is at such pressures and in finite magnetic
fields \cite{nofield} that we believe the incoherent
interplane transport is realized.  The
theoretical picture \cite{usadv,usmagic} is that 
in zero field the interplane hopping is just 
barely sufficient to stabilize a three 
dimensional Fermi liquid (were it not for
the superconducting transition).  For a
magnetic field applied along $c$, which
minimally disrupts interplane motion,
the superconductivity can be removed while
retaining interplane coherence, and the
behavior should be roughly Fermi liquid
like.  However,
for fields of sufficient strength
in other directions \cite{usmagic},
the magnetic field interferes with the
interchain coherence \cite{reallattice} and even moderate
fields in the $b$ direction 
completely remove interplane coherence.

\begin{figure}
\centerline{ \epsfxsize = 6in
\epsffile{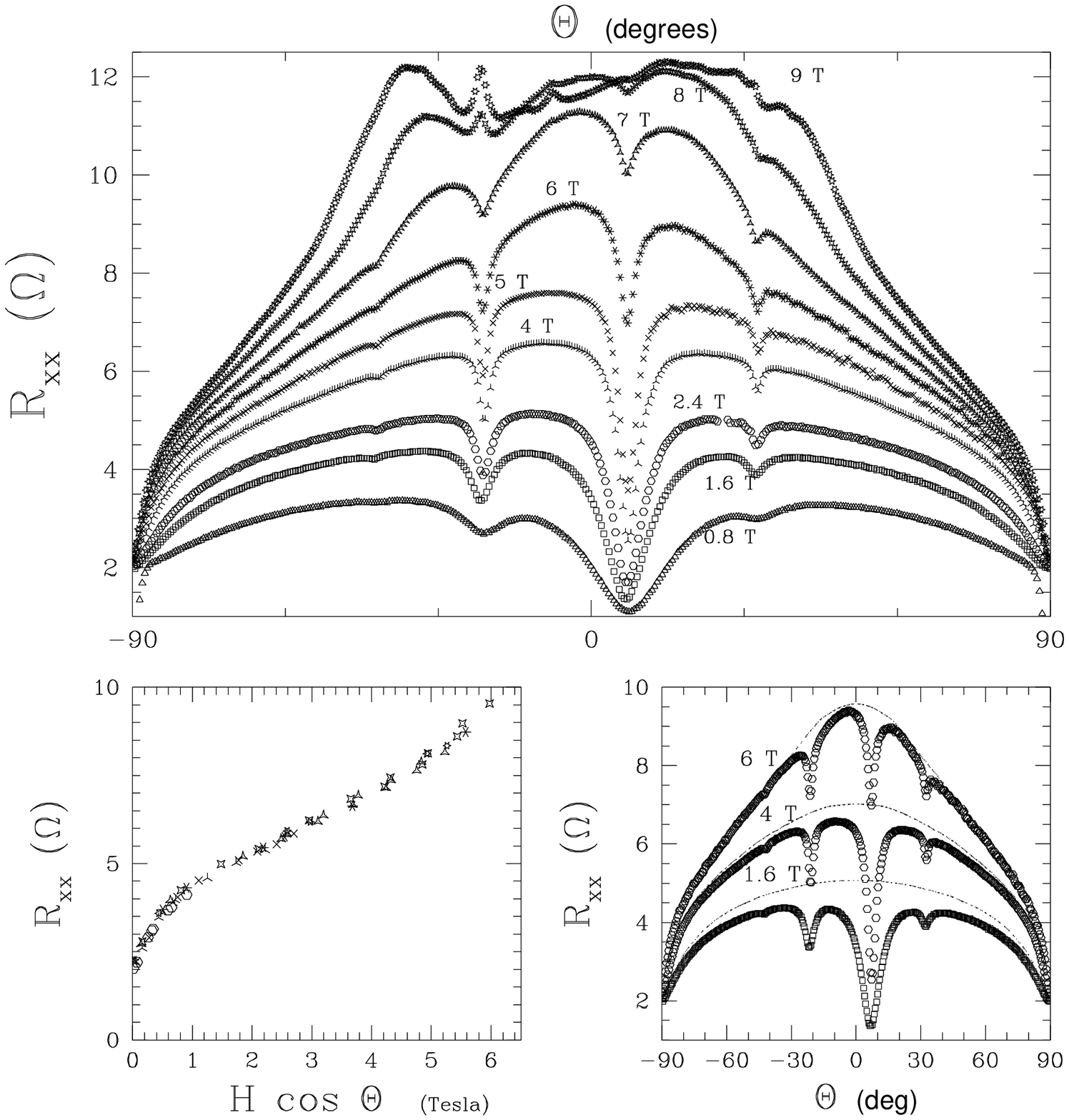}}
\caption{{\bf (A, Top-Center):} Magnetoresistance in the most conducting
lattice direction, $a$, as fields
of various strengths are rotated in the plane of the other two lattice
directions, i.e. from  $-b$ thru $c$ on to $b$.
$\Theta$ is measured from the perpendicular to the $b$ axis.
Data were taken at $10$ kilobar and $0.5$ Kelvin. 
{\bf (B, Bottom-Left):} Subset of data from A for field orientations
away from the ``magic angles''. Data are replotted as resistance versus field
strength out of the $ab$ plane.  Note the collapse 
onto a single scaling curve irrespective of field strength (see text).
{\bf (C, Bottom-Right):} Data for field
strengths of 1.6, 4 and 6 Tesla together with the expected magnetoresistance
from the scaling curve of part B. Deviations from scaling occur only
within a vicinity of the magic angles that decreases rapidly with field.}
\label{fig:xx}
\end{figure}

What then are the testable predictions of
this theoretical description?
The most natural experimental probes 
for coherence effects are 
low temperature
magnetotransport measurements, and it was 
a number of anomalies in these measurements that
first attracted our attention to the material.
Consider the data depicted in Figure \ref{fig:xx}.
So anisotropic a material should have a Fermi surface 
consisting of a pair of well separated sheets
and therefore the magnetoresistance
in the most conducting direction is expected in FLT 
to be very small and to saturate quickly \cite{victor}.  
Instead the the material displays an enormous, angle dependent
magnetoresistance for fields rotated in the $bc$ plane
and current in the $a$ direction.  Particularly
striking are the dip features which occur when
the magnetic field parallels a real space
lattice direction.  
%There have been various
%proposals put forward to explain the
%dips \cite{magicother};
In our proposal, the dips are naturally explained as places
where the magnetic field is ineffective in
disrupting interplane hopping and
coherence is not fully destroyed. Such
a picture has a number of qualitative predictions
such as the narrowing of the $c$ direction dip
roughly linearly with magnetic field and the
hierarchy of appearance of the dips (the $c$ dip 
must appear first as a function of field strength) \cite{usadv},
however, it is
away from the dips - where the quantum-classical
metal should be realized - that the strongest
experimental consequences of the confinement
of coherent motion to the planes should be 
apparent.  In that state it is impossible
to observe interference effects between
interplane histories and therefore the orbital
magnetoresistance contribution from the components
of the magnetic field lying in the $ab$ plane
{\em must vanish identically\/}.  Therefore,
the magnetoresistance data away from the magic angles
are also plotted in Fig. \ref{fig:xx},
not only versus angle \cite{zeeman}, but also versus the component
of the field perpendicular to the $ab$ plane.

Note the extent to which the data away from the magic angles
collapse onto a single curve, signaling the predicted 
independence of the magnetoresistance from
in-plane field strength. 
The data from within the magic angle dips, where some coherence
is present in our picture, are not at all independent of
in-plane field strength.
There is at present no other theoretical
explanation for this behavior
({\em e.g.\/} all the scenarios in \cite{magicother} fail to
exhibit the scaling observed in Fig. \ref{fig:xx}); at a minimum, it implies 
that
away from the dips, and only away from the dips,
the effect on the resistivity in this direction of interference
effects due to fields in the $ab$ plane is nearly zero.
Since there might be reasons for this
other than the incoherence of interplane hopping, 
let us consider further possible probes of the
coherence of interplane hopping.  

\begin{figure}
\centerline{ \epsfxsize = 6in
\epsffile{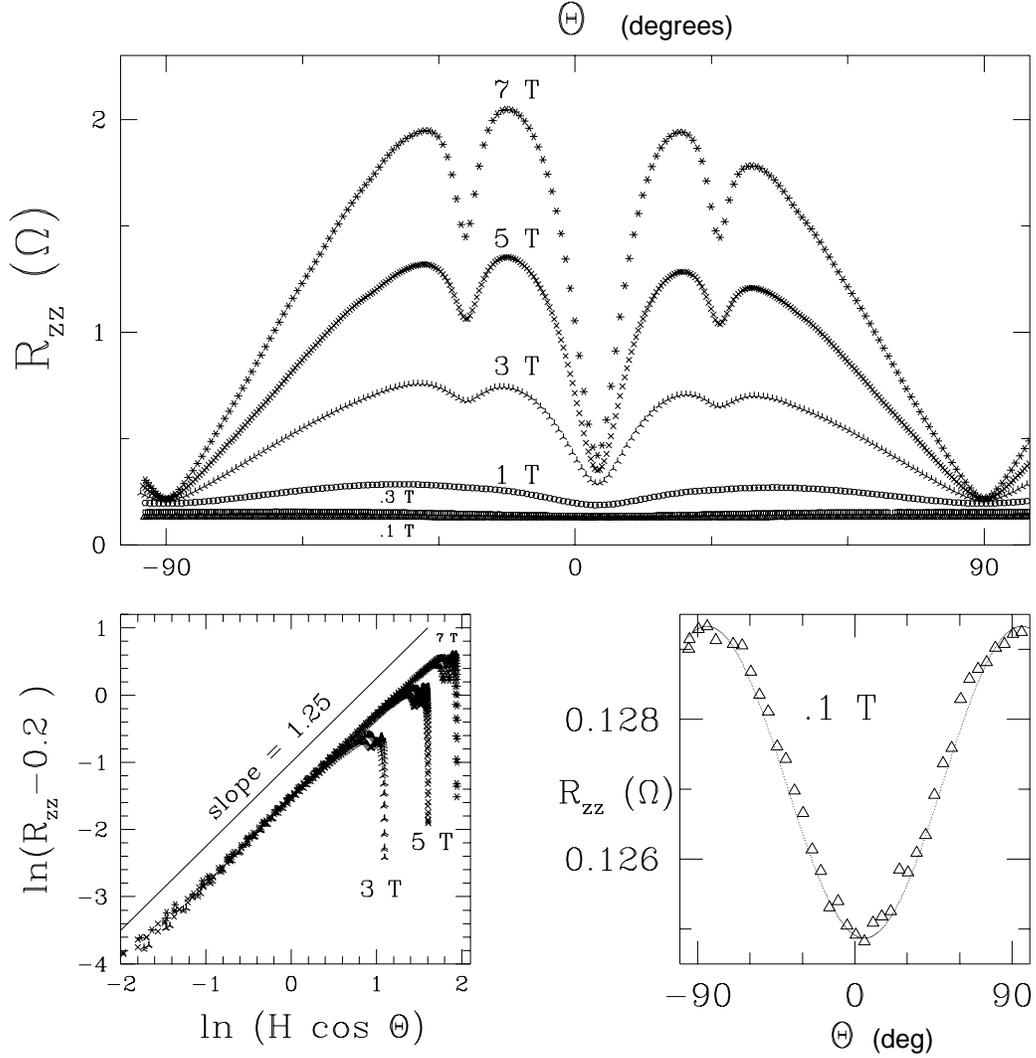}}
\caption{{\bf (A, Top-Center):} Magnetoresistance perpendicular to 
the $ab$ plane as fields of various strengths are rotated as in Figure 1.
Data were taken at $10$ kilobar and $1.3$ Kelvin.
{\bf (B, Bottom-Left):} Data from A for
3, 5, and 7 Tesla plotted as natural logarithm of deviation from a reference 
value versus natural logarithm of magnetic field strength perpendicular to
the $ab$ plane.  Note that, away from the ``magic angles'', the data
exhibit power law dependence on only one component of the magnetic field:
$\Delta R \propto (H \cos \Theta)^p$, $p \sim 1.25$ (see text).
{\bf (C, Bottom-Right):} Weak field ($0.1$ Tesla) magnetoresistance for $bc$ 
rotation.
Dotted line is a fit to the data of the form
$R_0 + \alpha |H \times \hat{c}|^2$, the semi-classical prediction.}
\label{fig:zz}
\end{figure}

Clearly, the most
natural quantity to consider in probing
$c$ axis coherence is the conductivity in 
this direction.  In a simple FLT model with an 
isotropic scattering rate the inverse of the conductivity
is easily calculated and behaves as 
$R_0 * \left( 1 + \tau^2 
(e v_F H_b)^2 \right)$
where $H_b$ is the component of the magnetic
field lying in the $ab$ plane and perpendicular to $a$.
In low fields, we expect coherence and
our theory of the magnetoresistance predicts that
something like the above behavior should
be observed in the $c$ direction resistivity,
while in high fields one expects the magic angle dips 
and the resistivity 
away from the dips to be independent of the in-plane field strength.
Recent data of Chashechkina \cite{katya} are shown in
Figure \ref{fig:zz}.  
The crossover from the low field,
approximately FLT behavior to magic angle behavior
is striking.  
Again, away from the magic angle dips, the resistivity 
depends only on the out of plane
component of the field, which is now the field
nearly parallel to the current!
At a minimum, the data require that
away from the dips, and only away from the dips,
the effect of interference effects, due to fields {\em in\/} the $ab$ plane,
on the resistivity {\em out\/} of the $ab$ plane,
is nearly zero.
Again, there is no theoretical proposal
other than the incoherence of the interplane hopping
which can account for the observed effect, especially
given the qualitative agreement between the low field data and
the expected behavior of a relatively clean Fermi liquid.

Danner \cite{guynfl} has also examined another 
interference effect \cite{guybatman} which is sensitive to the
coherence of $c$ axis transport and the existence of
three dimensional Fermi surface in (TMTSF)$_2$PF$_6$.
The results are again naturally explained by
the presence of coherence in fields whose projection
along certain real space lattice vectors is sufficiently
small, but the {\em total absence} of interplane
coherence for other fields.  

It is, of course, impossible to demonstrate the
total absence of coherence experimentally:
one can only show that various measured quantities
are consistent with the absence of interplane coherence,
i.e. they exhibit no signs of interference effects between
histories involving interplane motion.
Consequently, the above results cannot be taken
to {\em prove} the incoherence of $c$-axis electronic
motion; however, it is truly remarkable that
three different measures of interplane coherence
show the complete absence
of such interference effects. The magnetoresistance out
of the $ab$ plane is particularly striking since
it most directly probes the $c$ axis charge transport
and, while its low field behavior fits well into
the expectations of FLT,
its behavior in high fields is
totally anomalous from the FLT point of view.

\begin{figure}
\centerline{ \epsfxsize = 6in
\epsffile{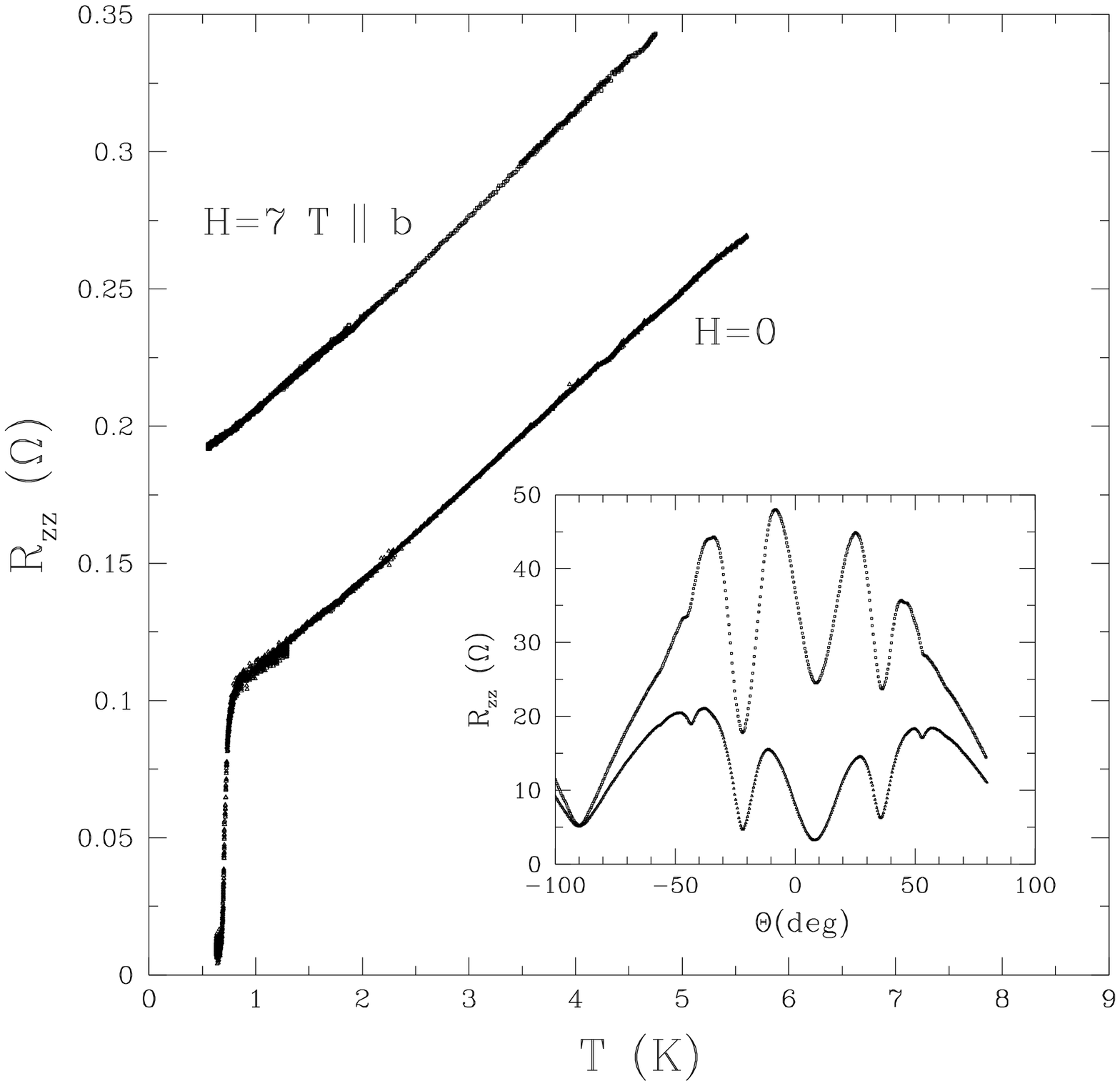}}
\vspace*{1.0in}
\caption{Temperature dependence of the resistivity perpendicular
to the $ab$ plane both in the absence of a magnetic field
and in a field of $7$ Tesla along $b$ (the latter condition is in the
incoherent phase). Data taken at $10$ kilobar applied 
pressure (see text). {\bf Inset:} Magnetoresistance perpendicular
to the $ab$ plane for $bc$ field rotations as in Figure 2 but at
$8.2$ kilobar and $50$ millikelvin.}
\label{fig:lowT}
\end{figure}

If one accepts that there is no coherent
interplane motion, is there any other explanation
besides our proposal of relevant interplane hopping which has been
driven incoherent by in-plane interaction effects?
Clearly,
if the hopping were irrelevant, {\em i.e.\/} the effective low energy
theory describing the system had no out of plane hopping,
the unimportance of magnetic fields in the
$ab$ plane would follow naturally. What would not
make sense, however, would be the strong angular dependence
of the $c$ direction magnetoresistance (particularly the dips)
or even the mere existence of
a substantial, low temperature $c$ axis conductivity.
Further, if the low energy theory has no $c$ axis
hopping of charge carriers, then the resistivity
must diverge as $T \rightarrow 0$. The experimental
behavior in the incoherent phase is shown in 
Fig. \ref{fig:lowT} and there is no evidence for
such behavior down to below a Kelvin.

Moreover, measurements 
at 50 mK (Fig. \ref{fig:lowT}) show no signs of a diverging resistivity.
It appears then that an explanation of the
unimportance of magnetic fields in the
$ab$ plane based on the absence of $c$
axis hopping is untenable.
If the hopping exists and is incoherent,
is this really a property of a low energy
fixed point or is it the result of
inelastic scattering or of disorder?  The essential features of
the magnetoresistance remain down to at least 
50 mK (Fig. \ref{fig:lowT}) so that an explanation
based on inelastic effects again
appears untenable. For many reasons, inelastic,
disorder scattering
is also unable to account for the loss
of coherence \cite{tdisordered} since: (1) the magic angles and scaling 
 are only observed in the
highest quality crystals; (2) the low field data 
are consistent
with a FLT type behavior with a scattering
rate of about $1K$, a rate too low to credibly explain
the incoherence of $c$ axis transport; (3) 
the magic angle dips in themselves demonstrate 
a strong dependence of the in-plane transport on the coherence of the
$c$ axis motion, a dependence which is
impossible if disorder dominates $c$ axis motion. 

\section{Conclusion}
It therefore appears to us that the
experimental situation
%, while not 
%rigorously unambigous, 
is remarkably compelling.
All indications are that in the limit of a pure system 
and zero temperature, (TMTSF)$_2$PF$_6$ 
exhibits a phase in which an applied magnetic
field \cite{nofield} not only destroys superconductivity,
but drives the system to a state of matter
characterized by finite, relevant interplane
electron hopping but the complete absence of
observable interference effects between histories
involving interplane motion. More succinctly, at zero temperature
there is a non-vanishing interplanar electron conductivity, but it is completely
incoherent.  This is to be contrasted with the coherent in-plane
transport. The state is, therefore,  
a non-Fermi liquid metal of a novel type, characterized by ``quantum''
in-plane, and ``classical'' inter-plane, transport: in short, 
a quantum-classical metal.

We gratefully acknowledge support from the following grants:
NSF grants MRSEC DMR-94-00362, 
DMR-9104873 (DGC), DMR-9626291 (PMC and EIC) and
DOE grant DE-FG02-90ER40542 (SPS).

%\end{multicols}


% \begin{table}
% \caption{}
% \label{}
% \end{tabular}
% \end{table}


\begin{references}
\bibitem{defs} Here $L$ is the system size and $Z$ is the overlap
between the bare electron and the nearly free quasiparticle
of FLT.
\bibitem{dimer}	The conduction band would be quarter filled
but at the temperatures considered here
the material is weakly dimerized along the most conducting
direction, resulting in a half-filled band.
\bibitem{nofield} The field required for incoherence appears
to be pressure dependent, decreasing with increasing pressure.
It is possible that the quantum-classical metal can occur
in this material
in zero magnetic field under sufficient pressure.
\bibitem{reallattice} Coherence may similarly survive if the
field is along a real space lattice direction with a component
out of the $ab$ plane and there is effective hopping in this
direction of sufficient strength. Hence the presence of 
additional magic angle dips.
\bibitem{zeeman} In the phase with incoherent interplane
hopping the orbital component of the magnetoresistance
cannot depend on the in plane components of the field;
however, there should be small Zeeman contributions
from these components even in the incoherent phase.
Empirically, this contribution is unobservable.
\bibitem{tdisordered} Disorder in the hopping between planes
is an even less tenable proposal than in-plane disorder;
as long as there is any spatially uniform
component to $t_{\perp}(x)$, its Fourier
transform has a delta function at zero momentum 
transfer with finite weight.
This finite amplitude for momentum conserving
single electron, interplane motion is generically
far more relevant than any momentum non-conserving
hopping process which might be invoked to account
for the loss of interplane coherence and should 
always dominate over such processes in the low energy limit.
\end{references}
\end{document}